\documentclass[aps,superscriptaddress,showpacs,reprint]{revtex4-1}

\usepackage[utf8]{inputenc}
\usepackage{amsmath,amsfonts,amssymb}
\usepackage{pifont}
\usepackage{rotating}
\usepackage{enumerate}
\usepackage{graphicx}
\usepackage[table,dvipsnames,svgnames,x11names]{xcolor}
\usepackage{soul}
\usepackage{subfigure}
\usepackage[normalem]{ulem}

\usepackage{multirow,bigstrut}
\newsavebox{\measurebox}

\usepackage{hyperref}
\definecolor{dark-red}{rgb}{0.9,0.15,0.15}
\definecolor{dark-blue}{rgb}{0.15,0.15,0.4}
\definecolor{dark2-blue}{rgb}{0.15,0.15,0.8}
\definecolor{medium-blue}{rgb}{0,0,0.5}
\hypersetup{
    colorlinks, linkcolor={black},
    citecolor={dark-blue}, urlcolor={medium-blue}
}
\begin{document}







\title{Spin-valve nature and giant coercivity of a ferrimagnetic spin semimetal Mn$_2$IrGa}

\author{Akhilesh Kumar Patel}
\thanks{These authors contributed equally to this work.}
\affiliation{Department of Physics, Indian Institute of Technology Bombay, Mumbai 400076, India}
\affiliation{Research Centre for Magnetic and Spintronic Materials, National Institute for Materials Science, Tsukuba, Ibaraki 305 0047, Japan}

\author{Y. Venkateswara}
\thanks{These authors contributed equally to this work.}
\affiliation{Department of Physics, Indian Institute of Technology Bombay, Mumbai 400076, India}
\affiliation{Spectroscopic Investigations of Novel Systems Laboratory, Department of Physics, Indian Institute of Technology Kanpur, Kanpur 208016, India}

\author{S. Shanmukharao Samatham}
\affiliation{Department of Physics, Chaitanya Bharathi Institute of Technology, Gandipet, Hyderabad 500 075, India}


\author{Archana Lakhani}
\affiliation{UGC-DAE Consortium for Scientific Research, University Campus, Indore 452001 Madhya Pradesh, India}

\author{Jayita Nayak}
\affiliation{Spectroscopic Investigations of Novel Systems Laboratory, Department of Physics, Indian Institute of Technology Kanpur, Kanpur 208016, India}

\author{K. G. Suresh}
\email{suresh@phy.iitb.ac.in}
\affiliation{Department of Physics, Indian Institute of Technology Bombay, Mumbai 400076, India}

\author{Aftab Alam}
\email{aftab@phy.iitb.ac.in}
\affiliation{Department of Physics, Indian Institute of Technology Bombay, Mumbai 400076, India}

\begin{abstract}
Spin semimetals are amongst the most recently discovered new class of spintronic materials, which exhibit a band gap in one spin channel and semimetallic feature in the other, thus facilitating tunable spin transport.	
Here, we report Mn$_2$IrGa to be a candidate material for spin semimetal along with giant coercivity and spin-valve characteristics using a combined experimental and theoretical study. The alloy crystallizes in an inverse Heusler structure (without any martensitic transition) with a para- to ferri-magnetic transition at $T_\mathrm{C} \sim$ 243 K. It shows a giant coercive field of about 8.5 kOe (at 2 K). The negative temperature coefficient, relatively low magnitude and weak temperture dependance of electrical resistivity suggest the semimetallic character of the alloy. This is further supported by our specific heat measurement. Magnetoresistance (MR) confirms an irreversible nature (with its magnitude $\sim$1\%) along with a change of sign across the magnetic transition indicating the potentiality of Mn$_2$IrGa in magnetic switching applications. In addition, asymmetric nature of MR in the positive and negative field cycles is indicative of spin-valve characteristics. Our ab-initio calculations confirm the inverse Heusler structure with ferrimagnetic ordering to be the lowest energy state, with a saturation magnetization of 2 $\mu_\mathrm{B}$. $<100>$ is found to be the easy magnetic axis with considerable magneto-crystalline anisotropy energy. A large positive Berry flux at/around $\Gamma$ point gives rise to an appreciable anomalous Hall conductivity ($\sim$-180 S/cm).

\end{abstract}


\date{\today}
\maketitle

\textbf{\it Introduction}: In the last two decades, the search for promising energy efficient multifunctional materials has gained tremendous momentum. Of these, new classes of magnetic materials such as spin gapless semiconductors, bipolar magnetic semiconductors, fully compensated ferrimagnetic and spin semimetals play a crucial role in taking the spintronic research to a different height.\cite{TGraf-SimpleRulesHeuslerAlloys,Ouardi-Mn2CoAl-SGS-PRL} Heusler alloys (HA) are one of the potential classes of materials, which host all of the above properties in different materials. They have also gained lot of attention due to their high magnetic ordering temperature (either $T_C$ or $T_N$), flexibility of doping/alloying enabling a wide platform for tunable band structure engineering. However, many of these alloys suffer from anti-site disorder, which often hinders their potential for application. HAs are mostly composed of two or more $3d$ transition elements. One of the ways to minimize the anti-site disorder is to replace one of the $3d$ transition elements by a $4d$ or $5d$ element. Some of the $4d$ based HAs reported for spintronics applications  are CoRhMnGe,\cite{CoRhMnGe-Deepika-exp-theory-PRB} CoRuMnSi,\cite{CoRuMnSi-Venkateswara-exp-theory-JMMM} Fe$_2$RhSi,\cite{Fe2RhSi-Venkateswara-exp-theory-PRB} VNbRuAl\cite{VNbRuAl-Nag-exp-theory-PRB} etc. There are only a few $5d$ based HAs that are experimentally reported to crystallize in cubic structure  e.g. Fe$_2$IrSi\cite{Fe2IrSi-Krishnamurthy-exp-JAP,Fe2IrSi-Lalrinkima-theory-PCCP}, LiGa$_2$Ir\cite{LiGa2Ir-Gornicka-exp-theory-SR} and Ru$_2$TaAl,\cite{Ru2TaAl-Tseng-exp-theory-PRB} though there are several theoretical reports such as Ir$_2$ScGa\cite{Ir2ScGa-Murat-theory-MTC}, Ir$_2$YSi (Y=Sc,Ti,V, Mn, Fe, Co, and Ni)\cite{Ir2YSi-Prakash-theory-AIPA}, Ir$_2$LuSb\cite{Ir2LuSb-Samia-theory-EM}, Ir$_2$ScAl\cite{Ir2ScAl-Arikan-theory-JKPS}, TiZrIrZ (Z=Al, Ga or In)\cite{TiZrIrZ-Muhammad-theory-TSF}, CoCrIrSi\cite{CoCrIrSi-Hoat-theory-CP}, ZnCdIrMn\cite{ZnCdIrMn-Han-theory-RP}, Zn$_2$IrMn\cite{Zn2IrMn-Han-theory-JMMM}, CoIrMnAl\cite{CoIrMnAl-Monma-theory-JAC}, Ru$_2$TaGa\cite{Ru2TaGa-Khandy-theory-JPCS}, Mn$_2$YZ (Y=Ta, W, Os, Ir, Pt, Au; Z=Ga,In).\cite{Mn2YZ-Wollmann-theory-PRB,Mn2YGa-Tayeb-theory-CJP,Mn2YGa-LiFan-theory-JMMM} Tetragonal phase of some of these alloys, e.g. Mn$_2$YZ, acquire high magnetic ordering temperature and show large exchange bias, spin-orbit coupling (SOC), anti-skyrmionic nature, giant magnetocaloric and large magnetoresistance (MR).\cite{Mn2PtGa-Nayak-exp-PRL,Mn2PtSn-Nayak-exp-Nature,Mn2RhSn-Meshcheriakova-exp-PRL}

Mn$_2$IrGa is an interesting system which is reported to stabilize in both cubic (\textit{F-43m}, \#219) and tetragonal (\textit{I-4m2}, \#119) structures.\cite{Mn2YZ-Wollmann-theory-PRB} In the cubic phase, the calculated equilibrium lattice constant is 5.97 \AA~and the net magnetization ($m$) is 2 $\mu_\mathrm{B}$/f.u. While, tetragonal phase shows low $m$ and large magneto-crystalline anisotropy, thus promising for spin-transfer magnetization-switching applications. First principles calculations by Hellal \textit{et al}.\cite{Mn2YGa-Tayeb-theory-CJP} reported ferromagnetic transition temperatures ($T_C$) of 368 K and 244 K for tetragonal and cubic phases respectively. As such, although there exist some theoretical studies on Mn$_2$IrGa, the experimental investigation on the structural, magnetic and electrical transport of this alloy is lacking. Specially, exploring its potential for spintronics related applications is highly desired.

In this letter, we report the structural, magnetic and electrical transport properties of \textcolor{black}{polycrystalline Mn$_2$IrGa} using a combined experimental and theoretical study. Mn$_2$IrGa crystallizes in an inverse cubic Heusler structure \textcolor{black}{with no signature of martensitic transition throughout the temperature}. Magnetic measurements reveal ferrimagnetic nature with giant coercive field ($\sim$8.5 kOe) below $T_\mathrm{C}$, \textcolor{black}{possibly due to strong SOC induced by Ir}. Zero-field-cooled (ZFC) and field-cooled (FC) T-dependent magnetization(M) shows large bifurcation. Real part of AC-susceptibility ($\chi_\mathrm{AC}'$) exhibits a sharp peak at/around 246 K and a broad peak at 25 K. Interestingly, the position of sharp peak is independent of frequency, indicating the long-range magnetic ordering, while the broad peak suggests the short range magnetic ordering at low T. Transport measurements reveal a semimetallic nature of Mn$_2$IrGa. MR data indicates giant hysteresis with asymmetric nature, indicating spin valve behavior. Specific heat data yields a moderate density of states (DOS) at the Fermi level along with a kink at $\sim$243 K indicating a possible transition. Ab-initio simulation predicts a ferrimagnetic spin semimetal behavior with a net moment of 2.0 $\mu_\mathrm{B}$. $<100>$ is found to be the easy magnetization axis with large  magnetocrystalline anisotropic energy. Simulated Berry curvature shows few hot spots in the Brillouin zone yielding a reasonably high anomalous Hall conductivity (-180 S/cm).
\begin{figure}[t]
\centering
\includegraphics[width=1\linewidth]{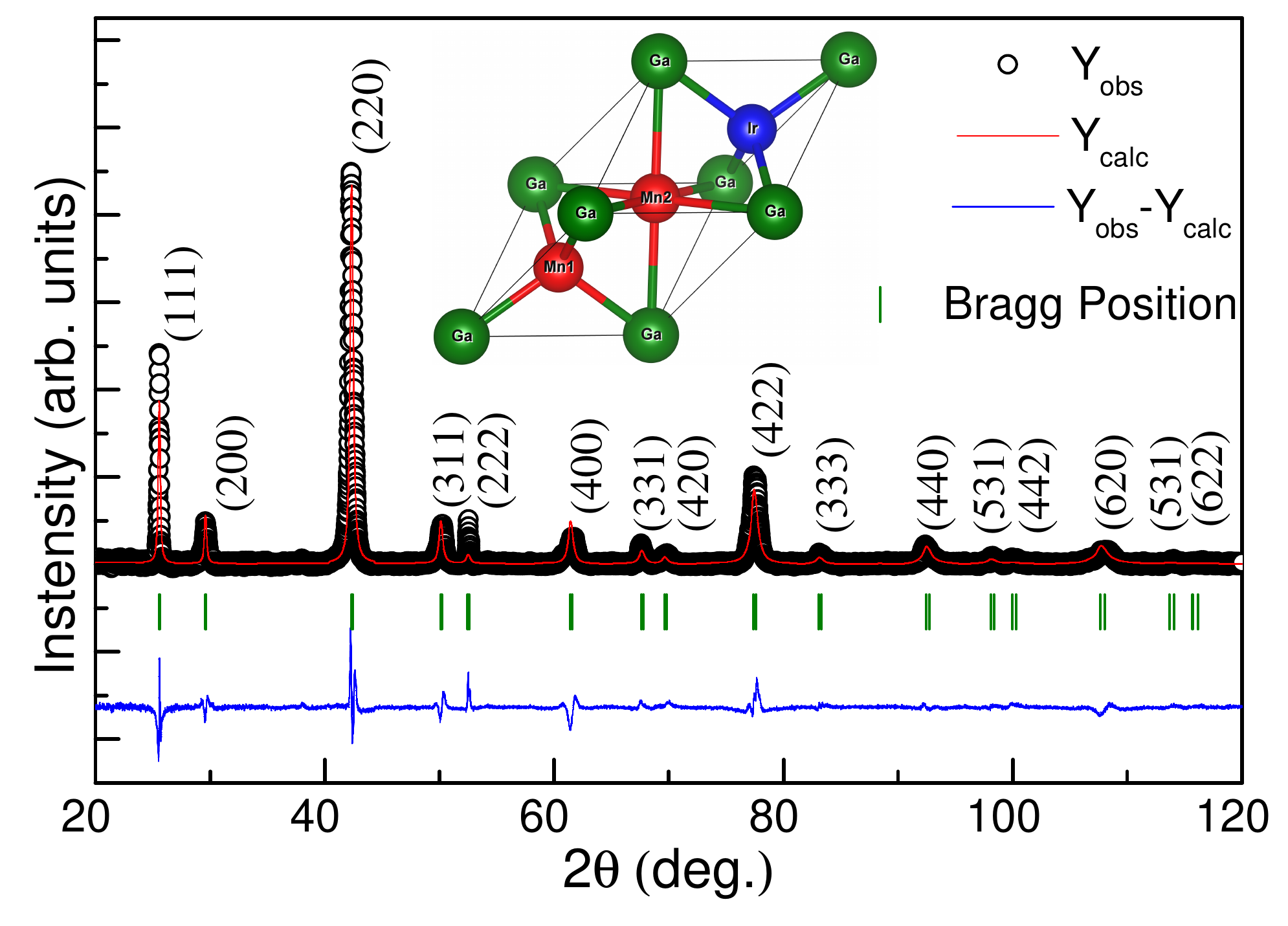}
\caption{Room temperature XRD pattern of Mn$_2$IrGa along with Rietveld refinement in Config. 1 whose structure is shown in the inset. }
\label{fig:XRD-Mn2IrGa}
\end{figure}

\textbf{\it Methods}: Polycrystalline Mn$_2$IrGa was prepared by arc-melting method with constituent elements of 4N purity. Further experimental details are given in Sec. A of the supplementary material (SM)\cite{sm} (see also Refs. \cite{FullProf-suite-cite, PhysRevB.99.174410, FEM-Simonson-PRB,PhysRevLett.109.246601,YVenkateswara-FeRhCrSi-PRB} therein)
First-principles calculations are performed using full potential linearised augmented plane- wave (FLAPW) method as implemented in FLEUR code.\cite{FleurRef1-PRB,FleurRef2-PRB,FleurRef3-PRB,FleurRef4-PRB} Other computational details are provided in Sec. B of SM.\cite{sm} Mn$_2$IrGa belongs to a full Heusler alloy (X$_2$YZ) where Z is a main group element.\cite{Simple-rules-TGraf-PSSC,Fe2RhSi-Venkateswara-exp-theory-PRB}  There exists two non-degenerate crystal configurations for this alloy, inverse Heusler structure (Config. I) and normal Heusler structure (Config. II). In the former,  the X atoms (here Mn) occupy two distinct Wyckoff sites namely the tetrahedral (4c) and the octahedral (4b) sites, while in the later, X atoms occupy the tetrahedral and Y occupy the octahedral sites.\cite{Fe2RhSi-Venkateswara-exp-theory-PRB}

\textbf{\it Results and Discussion}: Figure \ref{fig:XRD-Mn2IrGa}(a) shows the room temperature X-ray diffraction (XRD) pattern of Mn$_2$IrGa along with its Rietveld refined data. It crystallizes in the fcc structure (space group  F$\bar{4}$3m) with a lattice parameter of 6.03 \AA. The best fit is achieved for the inverse Heusler structure with Ga at 4a, Mn at 4b (say Mn1) and 4c (say Mn2), and Ir at 4d site, as shown in the inset of Fig. \ref{fig:XRD-Mn2IrGa}. Note that, it is only the minor (222) peak which has not fitted  well due to some texture along this direction. To further confirm the structural details, a high resolution transmission electron microscopy (HR-TEM) measurement is carried out (see Sec. C of SM\cite{sm}). The estimated d-spacing agrees fairly well with those obtained using XRD.

\begin{figure}[t]
	\centering
	\includegraphics[width=1\linewidth]{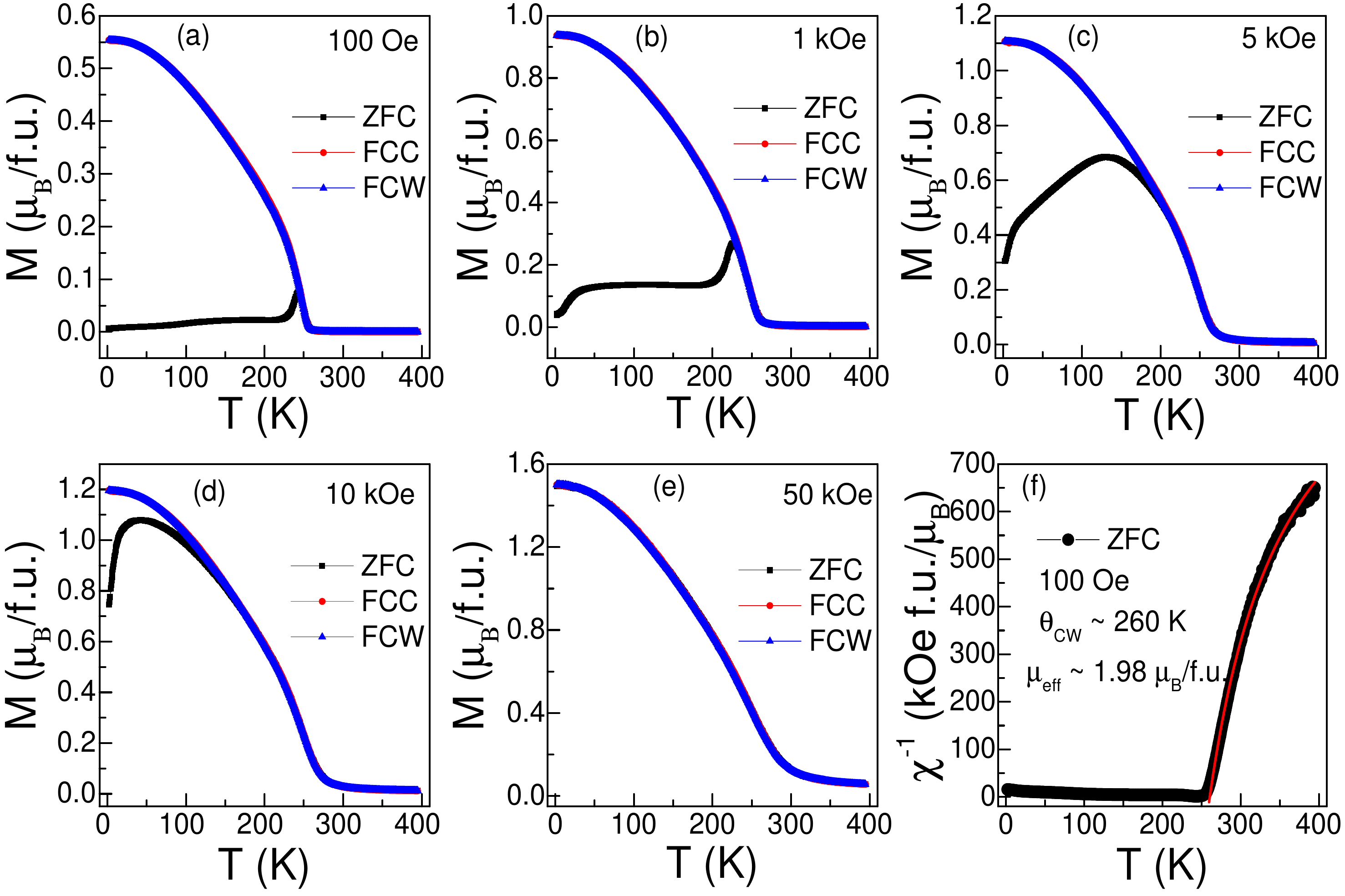}
	\caption{For Mn$_2$IrGa, $M$ vs. $T$ at various applied fields (a) 100 Oe, (b) 1 kOe, (c) 5 kOe, (d) 10 kOe and (e) 50 kOe. (f) Curie-Weiss fitting at 100 Oe.}
	\label{fig:M-Mn2IrGa}
\end{figure}

Figures \ref{fig:M-Mn2IrGa}(a)-(e) show the $T$-dependence of magnetization ($M$) at five different magnetic fields (H). Clearly, the field cooled cooling (FCC) and field cooled warming (FCW) curves coincide with each other, indicating the absence of first-order magnetic phase transition. Interestingly, a bifurcation between ZFC and FCC/FCW is noticed. A dimensionless quantity $\delta M = (M_{\mathrm{FCW}}-M_{\mathrm{ZFC}})/M_{\mathrm{ZFC}}$, quantifying the bifurcation is found to decrease with H and becomes zero at 50 kOe. Such bifurcation may originate from ferrimagnetic and/or spin glass behavior. Our theoretical simulations confirm the presence of ferrimagnetism and strong magneto-crystalline anisotropy (MCA) in the system. The bifurcation of ZFC and FCW can be explained with the help of MCA. Theoretically, $<$100$>$ crystal direction is found to be the easy axis while $<$110$>$ and $<$111$>$ are intermediate and hard axis respectively. In the polycrystalline sample, $<$100$>$ axes directions are oriented randomly. Hence, when the sample is cooled in the ZFC mode, the moments freeze randomly along $<$100$>$ directions. This phenomenon resembles a spin glass nature. The field applied for measuring the magnetization is not sufficient to reorient the moments along the field direction due to the anisotropy and hence leads to low magnetization compared to that in FCC mode. As the sample is cooled in the FCC mode, the moments freeze only along one of the preferred $<$100$>$ directions which leads to higher FCC magnetization. Figure \ref{fig:M-Mn2IrGa}(f) shows the Curie-Weiss fit for susceptibility using the expression $\chi^{-1}(T) =(T-\theta_\mathrm{CW})/(\chi_{0}(T-\theta_\mathrm{CW})+C)$. Here $\chi_{0}$, $C$ and $\theta_\mathrm{CW}$ are the $T$-independent part of $\chi$, Curie constant and Curie-Weiss temperature respectively. The effective magnetic moment can be estimated using $\mu_{\mathrm{eff}}=\sqrt{3Ck_\mathrm{B}/N_\mathrm{A}}$ where $N_A$ and $k_\mathrm{B}$ are the Avogadro's number and Boltzmann constants respectively. Curie-Weiss fitting yields $\theta_\mathrm{CW}\sim$ 260 K and $\mu_{\mathrm{eff}}$ $\sim$ 1.97 $\mu_\mathrm{B}$/f.u., which are in good agreement with other reports.\cite{Mn2YZ-Wollmann-theory-PRB} The real and imaginary part of susceptibility ($\chi$) indicating the magnetic and spin-glass transition is presented in Sec. D of SM.\cite{sm}

\begin{figure}[t]
	\centering
	\includegraphics[width=1\linewidth]{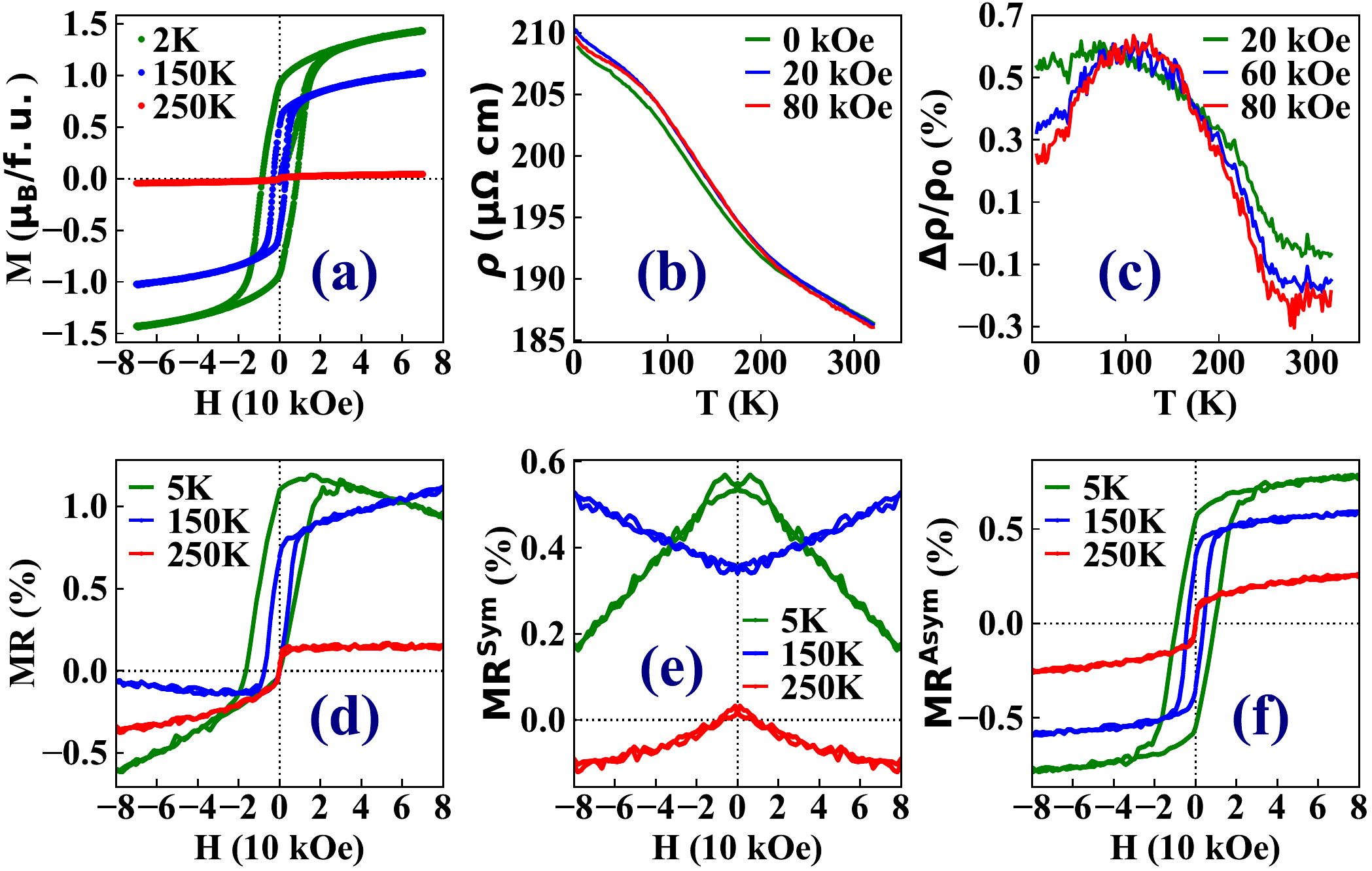}
	\caption{For Mn$_2$IrGa, (a) $M$ vs. $H$  at 2, 150 and 250 K (b) $T$-dependence of $\rho$, (c) $T$-dependence of magnetoresistance at different fields ($H$) (d) Isothermal magnetoresistance (MR) vs. field at 5, 150 and 250 K. (e,f) field dependance of symmetric and asymmetric parts of MR.}
	\label{fig:MHMR-Mn2IrGa}
\end{figure}

\textcolor{black}{
Figure \ref{fig:MHMR-Mn2IrGa}(a) shows the \textit{M-H} curves measured at few representative $T$. A finite hysteresis along with a non-saturating behavior of $M$ indicate the ferrimagnetic nature of the alloy. It shows a large coercive field (8.5 kOe) at 2 K which decreases with $T$. At 300 K, $M$ eases linearly with $H$ (not shown here) confirming the onset of paramagnetic behavior. Figure \ref{fig:MHMR-Mn2IrGa}(b) shows the electrical resistivity ($\rho$) vs. $T$ measured at few applied fields. It shows the negative temperature coefficient of resistivity, indicating the possibility of semimetal behavior. $\rho(T)$ undergoes slope change at two different temperatures $\sim100$ and $\sim240$ K. This categorises three different temperature regions. The crossover temperature ($\sim240$ K) is in accordance with the inferences made from susceptibility and specific heat (C($T$)) data (see Sec. D and G of SM\cite{sm} and footnote\cite{foot1}). The $H$-dependence of resistivity at different  $T$ is shown in Fig. S4 of SM\cite{sm}. $\rho(H)$ shows quite contrasting behavior in different $H$-range with varying $T$-values (see footnote\cite{foot2}).
}

\textcolor{black}{
Next, we estimated the $T$-dependance of magnetoresistance using MR(T)$ = [\rho(T,H)-\rho(T,H=0)]/\rho(T,H=0)$, as shown in Fig. \ref{fig:MHMR-Mn2IrGa}(c). MR($T$) clearly shows two turning points in accordance with the two slope changes in $\rho(T)$. It has a positive magnitude below $T_C$ while changes sign above $T_C$. In region I ($<$ 100 K), MR(T) increases with increasing $T$ and take a downhill in region II (100$<T<$240 K). The field ($H$) dependance of MR, as shown in Fig. \ref{fig:MHMR-Mn2IrGa}(d) at different $T$, shows giant hysteresis similar to $M$-$H$ curves, with a clear slope change at higher $H$. The symmetric and asymmetric parts of MR can be estimated as,
$\text{MR}^{type}(H) = [\text{MR}(H) \pm\text{MR}(-H)]/2 $, where $type$=Sym (Asym) takes positive (negative) sign on the right hand side. Note that due to the hysterisis, MR acquires symmetric and asymmetric components for both raising and lowering fields.
Figures \ref{fig:MHMR-Mn2IrGa}(e) and \ref{fig:MHMR-Mn2IrGa}(f) display the symmetric and asymmetric component of MR respectively. A more detailed MR data and its symmetric and asymmetric components at numerous $T$ are shown in Fig. S4 of SM\cite{sm}. The symmetric part of MR is negative for T$\ge$250 K. It is governed by the s-d scattering interaction with an almost linear variation with $H$. Below 250 K, although MR$^{Sym}$ remains +ve, it exhibits a crossover behavior with a sudden change in slope at 100 K, reflecting a similar change in $\rho$ vs. T behavior. Such variation of MR due to Lorentz contribution could arise if the product of cyclotron frequency and relaxation time is large.
The MR$^{\text{Asym}}$ has a clear hysteresis in regions I and II (i.e. 0-100 K and 100-240 K), whose magnitude decreases with increasing $T$, similar to that of $M-H$ curves (Fig. \ref{fig:MHMR-Mn2IrGa}(a)) while its hysteresis goes away for $T>240$. In fact, the $MR^{Asym}$ curves resemble very much with that of $M-H$ data indicating the electronic states that are responsible for magnetization, and contributing to its electrical transport.
}

Importantly, an asymmetric behavior of $\rho$ on positive and negative $H$-axis is clearly evident at all temperatures below 300 K (see Fig. S4 of SM). For all $T<T_\mathrm{C}$, Mn$_2$IrGa shows a two step asymmetric MR with hysteresis similar to \textit{M} vs. \textit{H} curves. This type of behavior generally arises in thin film heterostructures, where two ferromagnetic layers are separated by a nonmagnetic conducting layer \cite{nmat2983}, giving rise to the well-known spin-valve characteristics. Similar type of asymmetry is also reported in bulk Mn$_2$NiGa alloy.\cite{PhysRevLett.109.246601} Such behavior could be due to the minor anti-site disorder between a Ga and Mn atoms, which can be responsible for the formation of ferromagnetic (FM) nanoclusters (with parallel Mn spins) in a matrix of Mn$_2$IrGa bulk lattice having antiparallel Mn spin moments. The direction of Mn moments in the soft FM cluster reverses its direction with the application of field. This causes a rotation or tilt in the antiparallel Mn moments at the cluster-lattice interface, resulting in the observed asymmetry in MR.\cite{PhysRevLett.109.246601}


\begin{table}[t]
\centering
\caption{For Mn$_2$IrGa, theoretically optimized lattice parameters ($a_{eq}$), total and atom projected moments and the relative energies ($\Delta E$) of different magnetic ordering (ferrimagnetic(FiM), ferromagnetic(FM) and antiferromagnetic(AFM)) in the two structural configurations (I and II). Config. 1 with FiM ordering is set as the reference energy.}
\begin{tabular}{l c c c c c c }
		\hline \hline 	
		\multirow{3}{*}{Config. } & \multirow{3}{*}{$a_{eq}$ (\AA)} & \multicolumn{4}{c}{Moment ($\mu_B$)}  &  \multirow{3}{*}{$\Delta E (\mathrm{meV/atom})$} \\
		& & & & & \\
		&			 	&	4b		&	4c	&	4d		&	Total & 	\\ \hline
		&			&	$\mathrm{Mn1}$	&$\mathrm{Mn2}$	&	$\mathrm{Ir}$ & 	& \\
		
		I(FiM)  &   5.96 &		3.00			&	-1.41			&		0.22		& 2.00	& 	0.0 \\
		
		&		&					&				&			&		& 	\\
		&			&	$\mathrm{Ir}$	&$\mathrm{Mn1}$	&	$\mathrm{Mn2}$ & 	& \\		
		II(FM)	 & 6.12	&		0.69			&	3.30			&		3.30		& 7.74	& 170.7	\\ 
		&		&					&				&			&		& 	\\		
		II (AFM)	 & 6.09	&		0.00			&	-3.08			&		3.08		& 0.00	& 186.3	\\ 	
		\hline \hline
		
\end{tabular}
\label{tab:total_energies}
\end{table}

\begin{figure}[t]
\centering
\includegraphics[width=\linewidth]{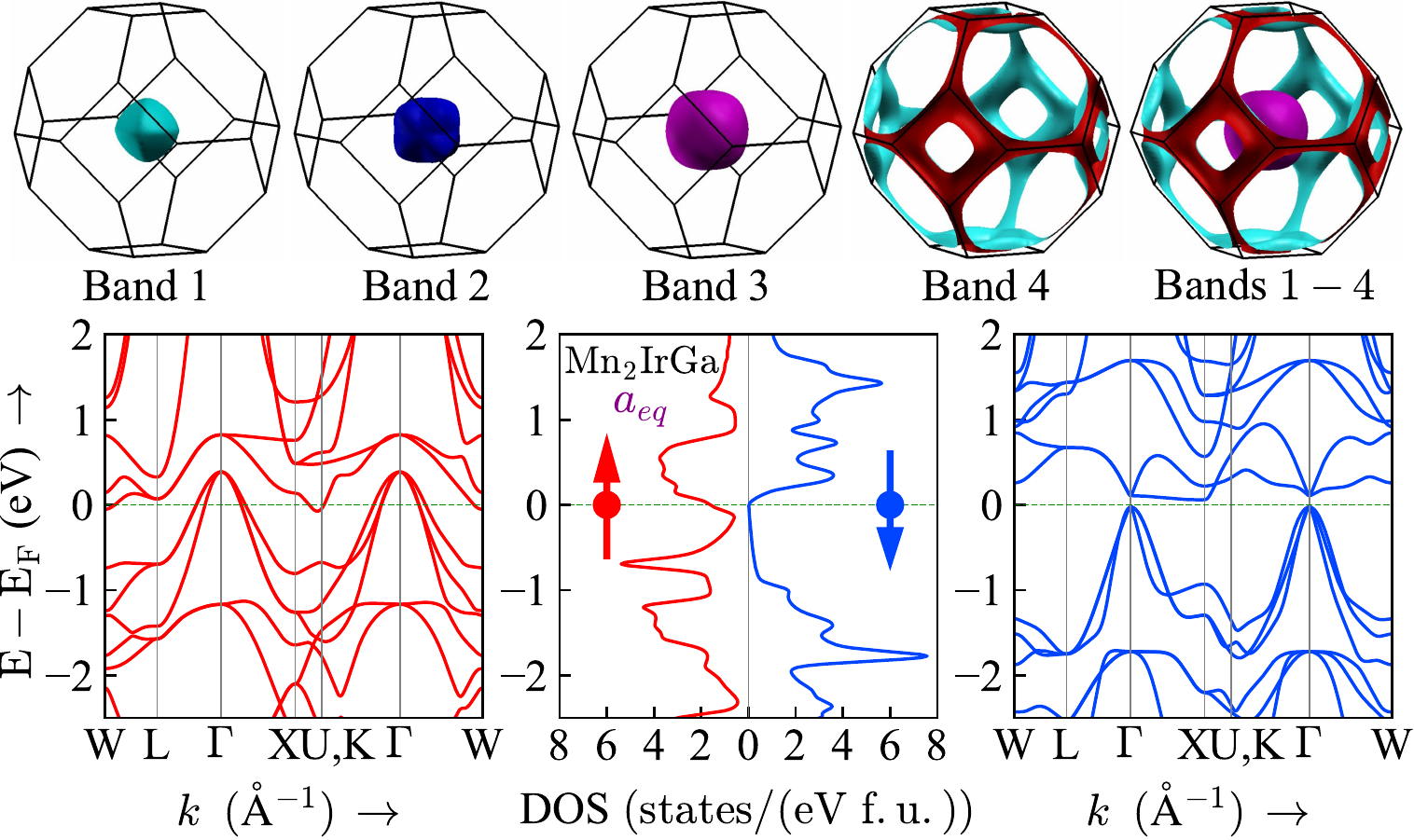}
\caption{(Bottom) Spin resolved band structure and density of states of Mn$_2$IrGa in its lowest energy Config. I (with FiM ordering). (Top) Fermi surfaces due to spin up bands \#1, 2, 3 and 4.}
\label{fig:DB-Mn2IrGa-nosoc}
\end{figure}

Ab-initio total energy calculations are performed for two different crystal configurations I, II (see Methods section) considering different magnetic arrangements (ferro-, antiferro- and ferri-magnetic). In case of Config. I, a unique ferrimagnetic (FiM) ordering got stabilized while for Config. II, two different magnetic ordering (FM and AFM) were realized with energy difference of 16 meV. FiM is the lowest energy magnetic ordering with the net moment of 2 $\mu_B$/f.u. Table \ref{tab:total_energies} shows the optimized lattice parameters, total and atom projected moments and the relative energies for different magnetic states in the two configurations.
The theoretically optimized lattice parameter for FiM is  $5.97$ \AA which matches fairly well with the experimental value. The local moments at the octahedral Mn (Mn1), tetrahedral Mn (Mn2) and Ir are 3.00, -1.41 and 0.22 $\mu_\mathrm{B}$ respectively for the FiM state.
Figure \ref{fig:DB-Mn2IrGa-nosoc} (lower panel) shows the spin polarized band structure and density of states for the lowest energy FiM state (the same for AFM and FM ordering is shown in Sec. H of  SM\cite{sm}). The spin down band shows a narrow band gap while the spin up band has an indirect overlap between valence and conduction bands. For spin up channel, the valence band involves three hole mediated metallic states while the conduction band contain two flat bands. The flat bands generally gives rise to hole and electron pockets near $E_F$. With increase in temperature, these pockets help to create/hold more charge carriers thereby dominating phonon contribution to electrical resistivity. Even though the valence band contains hole mediated metallic states, its contribution is restricted by several factors such as (i) high effective mass, (ii) trapping or scattering due to disorder in the lattice etc. The three bands crossing the $E_F$ forming the hole pockets around the $\Gamma$ point are labelled as `Band 1', `Band 2' and `Band 3'  whose Fermi surfaces are show in Fig. \ref{fig:DB-Mn2IrGa-nosoc} (top). `Band 4' gives rise to the electron pocket arising out of one of the conduction band crossing the $E_F$, whose Fermi surface is also shown in Fig. \ref{fig:DB-Mn2IrGa-nosoc}.

\begin{figure}[b]
\centering
\includegraphics[width=\linewidth]{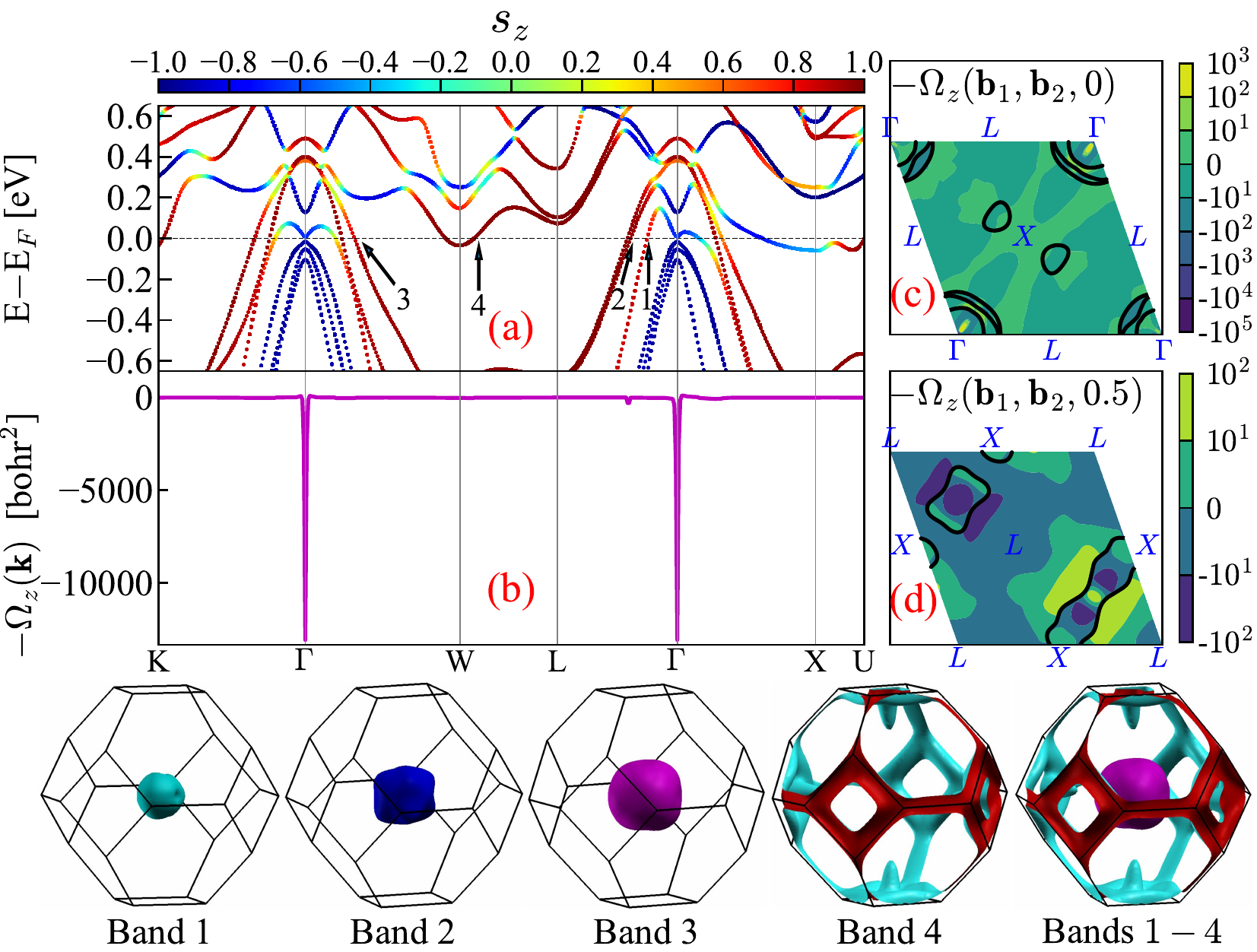}
\caption{For  Mn$_2$IrGa, (a) wannierized band structure including SOC, with $z-$component of spin moment shown by the color profile. 1, 2, 3 and 4 indicate the band \# crossing $E_F$ whose Fermi surfaces are shown at the bottom. (b) Berry curvature along the high symmetry $k-$paths. (c,d) 2D projected Berry curvature contours and Fermi lines (shown by black lines) on $\mathbf{b_3}=0$ and $\mathbf{b_3}=0.5$($\frac{2\pi}{a}$) planes.}
\label{fig:Mn2IrGa-bands+curv_z_soc}
\end{figure}

We further performed  the calculations including the effect of spin orbit coupling with different magnetization vector orientations $<$100$>$, $<$010$>$, $<$001$>$, $<$110$>$, $<$101$>$, $<$011$>$ and $<$111$>$. $<$100$>$ is found to be the easy axis, while $<$110$>$ and  $<$111$>$ are the intermediate and hard axes respectively. For cubic symmetry, the magnetic anisotropic energy can be expressed as\cite{Cullity-Graham-IMM}
\begin{equation}
E_{MAE}(\theta,\phi) = k_0 + k_1 (\alpha^2\beta^2 + \beta^2\gamma^2 + \gamma^2\alpha^2) + k_2 \alpha^2\beta^2\gamma^2\nonumber
\end{equation}  
where $\alpha$, $\beta$ and $\gamma$ are the direction cosines of magnetization axis with respect to the crystallographic axes. $k_1$ and $k_2$ are estimated using the eqns, $k_1=4(E^{<110>}-E^{<100>})$,  $k_2=9(E^{<111>}+3E^{<100>}-4E^{<110>})$ with the values $1.99\times10^5$ $J/m^3$, $5.20\times10^5$ $J/m^3$ respectively. Using these values of $k_1$ and $k_2$, $E_{MAE}$ is simulated as a function of $\theta$ and $\phi$, as shown in Sec. H of SM.\cite{sm}  This clearly confirms the anisotropic nature of the intrinsic magneto-crystalline energy. \textcolor{black}{Such large anisotropy can be responsible for narrow domain walls, which in turn causes hysteresis in the MR and magnetization isotherms. }

Next, we simulated the band structure including SOC considering the easy axis ($<$100$>$). Figure \ref{fig:Mn2IrGa-bands+curv_z_soc}(a) shows the Wannierized orbital projected band structure along with the $z-$component of spin moment ($s_z$).
Spin down character is clearly visible along $\Gamma$ to $X$ with relatively reduced band gap, while mixed spin character for bands 1 and 3 appear along several directions. Only band 2 retains its pure spin up character maintaining its Fermi surface unchanged (see `Band 2' Fermi surface in bottom figure). A slight pinched Fermi surface is noticed for bands 1, 3 and 4 due to the SOC mediated band splitting. Figure \ref{fig:Mn2IrGa-bands+curv_z_soc}(b) shows the intrinsic Berry curvature along the high symmetry $k$-paths, with strong spikes at $\Gamma$-point where the band crossing near $E_F$ occur.
Figure \ref{fig:Mn2IrGa-bands+curv_z_soc}(c,d) show the 2D projected Berry curvature contours on the $b_3=0$ and $b_3=\pi/a_{eq}$ planes. The black solid lines highlight the Fermi lines. Along with the large negative value, $-\Omega_z$ also has large positive values around the $\Gamma $ point (encircled like a toroid in Fig. S6 of SM\cite{sm}). As such, the Berry flux at/around $\Gamma$  resembles the magnetic flux in a bar magnet. The calculated intrinsic anomalous Hall conductivity is found to be $\sim -180$ S/cm.

\textbf{\it Summary}: We report Mn$_2$IrGa to be a potential candidate for the recently discovered ferrimagnetic spin semimetal (SSM). In SSM, one spin channel shows semimetallic behavior while the other has a narrow band gap. Apart from spin semimetallic feature, Mn$_2$IrGa also shows giant coercivity and spin-valve like characteristics. It crystallizes in an inverse Heusler structure with ferrimagnetic (FiM) ordering and a net magnetization of 2 $\mu_B$/f.u., \textcolor{black}{with no signature of martensitic transition.} Negative temperature coefficient of resistivity with a very weak variation with temperature indicate the semimetallic behavior of the alloy. Asymmetric nature of magneto-resistance with hysteresis and a change in sign across the transition temperature indicate its potential to be used in magnetic switching applications. \textit{Ab-initio} calculations confirm the SSM behavior with a unique FiM ordering. $<100>$ is simulated to be the easy magnetic axis with considerable anisotropy energy, \textcolor{black}{possibly causing narrow domain walls responsible for appreciable hysteresis in magnetization and MR.} Our simulation confirms a large intrinsic Berry curvature mediating a reasonably high anomalous Hall conductivity ($\sim$ -180 S/cm).

\textbf{\it Acknowledgments}: SSS acknowledges SERB for the financial support through Core Research Grant (CRG/2022/0007993). AA acknowledges DST-SERB (Grant No. CRG/2019/002050) for funding to support this research

\bibliographystyle{apsrev4-1}
\bibliography{bib}

\begin{thebibliography}{43}%
\makeatletter
\providecommand \@ifxundefined [1]{%
 \@ifx{#1\undefined}
}%
\providecommand \@ifnum [1]{%
 \ifnum #1\expandafter \@firstoftwo
 \else \expandafter \@secondoftwo
 \fi
}%
\providecommand \@ifx [1]{%
 \ifx #1\expandafter \@firstoftwo
 \else \expandafter \@secondoftwo
 \fi
}%
\providecommand \natexlab [1]{#1}%
\providecommand \enquote  [1]{``#1''}%
\providecommand \bibnamefont  [1]{#1}%
\providecommand \bibfnamefont [1]{#1}%
\providecommand \citenamefont [1]{#1}%
\providecommand \href@noop [0]{\@secondoftwo}%
\providecommand \href [0]{\begingroup \@sanitize@url \@href}%
\providecommand \@href[1]{\@@startlink{#1}\@@href}%
\providecommand \@@href[1]{\endgroup#1\@@endlink}%
\providecommand \@sanitize@url [0]{\catcode `\\12\catcode `\$12\catcode
  `\&12\catcode `\#12\catcode `\^12\catcode `\_12\catcode `\%12\relax}%
\providecommand \@@startlink[1]{}%
\providecommand \@@endlink[0]{}%
\providecommand \url  [0]{\begingroup\@sanitize@url \@url }%
\providecommand \@url [1]{\endgroup\@href {#1}{\urlprefix }}%
\providecommand \urlprefix  [0]{URL }%
\providecommand \Eprint [0]{\href }%
\providecommand \doibase [0]{http://dx.doi.org/}%
\providecommand \selectlanguage [0]{\@gobble}%
\providecommand \bibinfo  [0]{\@secondoftwo}%
\providecommand \bibfield  [0]{\@secondoftwo}%
\providecommand \translation [1]{[#1]}%
\providecommand \BibitemOpen [0]{}%
\providecommand \bibitemStop [0]{}%
\providecommand \bibitemNoStop [0]{.\EOS\space}%
\providecommand \EOS [0]{\spacefactor3000\relax}%
\providecommand \BibitemShut  [1]{\csname bibitem#1\endcsname}%
\let\auto@bib@innerbib\@empty
\bibitem [{\citenamefont {Graf}\ \emph
  {et~al.}(2011{\natexlab{a}})\citenamefont {Graf}, \citenamefont {Felser},\
  and\ \citenamefont {Parkin}}]{TGraf-SimpleRulesHeuslerAlloys}%
  \BibitemOpen
  \bibfield  {author} {\bibinfo {author} {\bibfnamefont {T.}~\bibnamefont
  {Graf}}, \bibinfo {author} {\bibfnamefont {C.}~\bibnamefont {Felser}}, \ and\
  \bibinfo {author} {\bibfnamefont {S.~S.}\ \bibnamefont {Parkin}},\ }\href
  {\doibase https://doi.org/10.1016/j.progsolidstchem.2011.02.001} {\bibfield
  {journal} {\bibinfo  {journal} {Progress in Solid State Chemistry}\ }\textbf
  {\bibinfo {volume} {39}},\ \bibinfo {pages} {1} (\bibinfo {year}
  {2011}{\natexlab{a}})}\BibitemShut {NoStop}%
\bibitem [{\citenamefont {Ouardi}\ \emph {et~al.}(2013)\citenamefont {Ouardi},
  \citenamefont {Fecher}, \citenamefont {Felser},\ and\ \citenamefont
  {K\"ubler}}]{Ouardi-Mn2CoAl-SGS-PRL}%
  \BibitemOpen
  \bibfield  {author} {\bibinfo {author} {\bibfnamefont {S.}~\bibnamefont
  {Ouardi}}, \bibinfo {author} {\bibfnamefont {G.~H.}\ \bibnamefont {Fecher}},
  \bibinfo {author} {\bibfnamefont {C.}~\bibnamefont {Felser}}, \ and\ \bibinfo
  {author} {\bibfnamefont {J.}~\bibnamefont {K\"ubler}},\ }\href {\doibase
  10.1103/PhysRevLett.110.100401} {\bibfield  {journal} {\bibinfo  {journal}
  {Phys. Rev. Lett.}\ }\textbf {\bibinfo {volume} {110}},\ \bibinfo {pages}
  {100401} (\bibinfo {year} {2013})}\BibitemShut {NoStop}%
\bibitem [{\citenamefont {Rani}\ \emph {et~al.}(2017)\citenamefont {Rani},
  \citenamefont {Enamullah}, \citenamefont {Suresh}, \citenamefont {Yadav},
  \citenamefont {Jha}, \citenamefont {Bhattacharyya}, \citenamefont {Varma},\
  and\ \citenamefont {Alam}}]{CoRhMnGe-Deepika-exp-theory-PRB}%
  \BibitemOpen
  \bibfield  {author} {\bibinfo {author} {\bibfnamefont {D.}~\bibnamefont
  {Rani}}, \bibinfo {author} {\bibnamefont {Enamullah}}, \bibinfo {author}
  {\bibfnamefont {K.~G.}\ \bibnamefont {Suresh}}, \bibinfo {author}
  {\bibfnamefont {A.~K.}\ \bibnamefont {Yadav}}, \bibinfo {author}
  {\bibfnamefont {S.~N.}\ \bibnamefont {Jha}}, \bibinfo {author} {\bibfnamefont
  {D.}~\bibnamefont {Bhattacharyya}}, \bibinfo {author} {\bibfnamefont {M.~R.}\
  \bibnamefont {Varma}}, \ and\ \bibinfo {author} {\bibfnamefont
  {A.}~\bibnamefont {Alam}},\ }\href {\doibase 10.1103/PhysRevB.96.184404}
  {\bibfield  {journal} {\bibinfo  {journal} {Phys. Rev. B}\ }\textbf {\bibinfo
  {volume} {96}},\ \bibinfo {pages} {184404} (\bibinfo {year}
  {2017})}\BibitemShut {NoStop}%
\bibitem [{\citenamefont {Venkateswara}\ \emph {et~al.}(2020)\citenamefont
  {Venkateswara}, \citenamefont {Rani}, \citenamefont {Suresh},\ and\
  \citenamefont {Alam}}]{CoRuMnSi-Venkateswara-exp-theory-JMMM}%
  \BibitemOpen
  \bibfield  {author} {\bibinfo {author} {\bibfnamefont {Y.}~\bibnamefont
  {Venkateswara}}, \bibinfo {author} {\bibfnamefont {D.}~\bibnamefont {Rani}},
  \bibinfo {author} {\bibfnamefont {K.}~\bibnamefont {Suresh}}, \ and\ \bibinfo
  {author} {\bibfnamefont {A.}~\bibnamefont {Alam}},\ }\href {\doibase
  https://doi.org/10.1016/j.jmmm.2020.166536} {\bibfield  {journal} {\bibinfo
  {journal} {Journal of Magnetism and Magnetic Materials}\ }\textbf {\bibinfo
  {volume} {502}},\ \bibinfo {pages} {166536} (\bibinfo {year}
  {2020})}\BibitemShut {NoStop}%
\bibitem [{\citenamefont {Venkateswara}\ \emph {et~al.}(2021)\citenamefont
  {Venkateswara}, \citenamefont {Samatham}, \citenamefont {Patel},
  \citenamefont {Babu}, \citenamefont {Varma}, \citenamefont {Suresh},\ and\
  \citenamefont {Alam}}]{Fe2RhSi-Venkateswara-exp-theory-PRB}%
  \BibitemOpen
  \bibfield  {author} {\bibinfo {author} {\bibfnamefont {Y.}~\bibnamefont
  {Venkateswara}}, \bibinfo {author} {\bibfnamefont {S.~S.}\ \bibnamefont
  {Samatham}}, \bibinfo {author} {\bibfnamefont {A.~K.}\ \bibnamefont {Patel}},
  \bibinfo {author} {\bibfnamefont {P.~D.}\ \bibnamefont {Babu}}, \bibinfo
  {author} {\bibfnamefont {M.~R.}\ \bibnamefont {Varma}}, \bibinfo {author}
  {\bibfnamefont {K.~G.}\ \bibnamefont {Suresh}}, \ and\ \bibinfo {author}
  {\bibfnamefont {A.}~\bibnamefont {Alam}},\ }\href {\doibase
  10.1103/PhysRevB.104.094402} {\bibfield  {journal} {\bibinfo  {journal}
  {Phys. Rev. B}\ }\textbf {\bibinfo {volume} {104}},\ \bibinfo {pages}
  {094402} (\bibinfo {year} {2021})}\BibitemShut {NoStop}%
\bibitem [{\citenamefont {Nag}\ \emph {et~al.}(2021)\citenamefont {Nag},
  \citenamefont {Rani}, \citenamefont {Kangsabanik}, \citenamefont {Singh},
  \citenamefont {Venkatesh}, \citenamefont {Babu}, \citenamefont {Suresh},\
  and\ \citenamefont {Alam}}]{VNbRuAl-Nag-exp-theory-PRB}%
  \BibitemOpen
  \bibfield  {author} {\bibinfo {author} {\bibfnamefont {J.}~\bibnamefont
  {Nag}}, \bibinfo {author} {\bibfnamefont {D.}~\bibnamefont {Rani}}, \bibinfo
  {author} {\bibfnamefont {J.}~\bibnamefont {Kangsabanik}}, \bibinfo {author}
  {\bibfnamefont {D.}~\bibnamefont {Singh}}, \bibinfo {author} {\bibfnamefont
  {R.}~\bibnamefont {Venkatesh}}, \bibinfo {author} {\bibfnamefont {P.~D.}\
  \bibnamefont {Babu}}, \bibinfo {author} {\bibfnamefont {K.~G.}\ \bibnamefont
  {Suresh}}, \ and\ \bibinfo {author} {\bibfnamefont {A.}~\bibnamefont
  {Alam}},\ }\href {\doibase 10.1103/PhysRevB.104.134406} {\bibfield  {journal}
  {\bibinfo  {journal} {Phys. Rev. B}\ }\textbf {\bibinfo {volume} {104}},\
  \bibinfo {pages} {134406} (\bibinfo {year} {2021})}\BibitemShut {NoStop}%
\bibitem [{\citenamefont {Krishnamurthy}\ \emph {et~al.}(2003)\citenamefont
  {Krishnamurthy}, \citenamefont {Weston}, \citenamefont {Mankey},
  \citenamefont {Suzuki}, \citenamefont {Kawamura},\ and\ \citenamefont
  {Ishikawa}}]{Fe2IrSi-Krishnamurthy-exp-JAP}%
  \BibitemOpen
  \bibfield  {author} {\bibinfo {author} {\bibfnamefont {V.~V.}\ \bibnamefont
  {Krishnamurthy}}, \bibinfo {author} {\bibfnamefont {J.~L.}\ \bibnamefont
  {Weston}}, \bibinfo {author} {\bibfnamefont {G.~J.}\ \bibnamefont {Mankey}},
  \bibinfo {author} {\bibfnamefont {M.}~\bibnamefont {Suzuki}}, \bibinfo
  {author} {\bibfnamefont {N.}~\bibnamefont {Kawamura}}, \ and\ \bibinfo
  {author} {\bibfnamefont {T.}~\bibnamefont {Ishikawa}},\ }\href {\doibase
  10.1063/1.1558274} {\bibfield  {journal} {\bibinfo  {journal} {Journal of
  Applied Physics}\ }\textbf {\bibinfo {volume} {93}},\ \bibinfo {pages} {7981}
  (\bibinfo {year} {2003})}\BibitemShut {NoStop}%
\bibitem [{\citenamefont {Lalrinkima}\ \emph {et~al.}(2021)\citenamefont
  {Lalrinkima}, \citenamefont {Ekuma}, \citenamefont {Chibueze}, \citenamefont
  {Fomin}, \citenamefont {Malikov}, \citenamefont {Zadeng},\ and\ \citenamefont
  {Rai}}]{Fe2IrSi-Lalrinkima-theory-PCCP}%
  \BibitemOpen
  \bibfield  {author} {\bibinfo {author} {\bibnamefont {Lalrinkima}}, \bibinfo
  {author} {\bibfnamefont {C.~E.}\ \bibnamefont {Ekuma}}, \bibinfo {author}
  {\bibfnamefont {T.~C.}\ \bibnamefont {Chibueze}}, \bibinfo {author}
  {\bibfnamefont {L.~A.}\ \bibnamefont {Fomin}}, \bibinfo {author}
  {\bibfnamefont {I.~V.}\ \bibnamefont {Malikov}}, \bibinfo {author}
  {\bibfnamefont {L.}~\bibnamefont {Zadeng}}, \ and\ \bibinfo {author}
  {\bibfnamefont {D.~P.}\ \bibnamefont {Rai}},\ }\href {\doibase
  10.1039/D1CP00418B} {\bibfield  {journal} {\bibinfo  {journal} {Phys. Chem.
  Chem. Phys.}\ }\textbf {\bibinfo {volume} {23}},\ \bibinfo {pages} {11876}
  (\bibinfo {year} {2021})}\BibitemShut {NoStop}%
\bibitem [{\citenamefont {Górnicka}\ \emph {et~al.}(2021)\citenamefont
  {Górnicka}, \citenamefont {Kuderowicz}, \citenamefont {Winiarski},
  \citenamefont {Wiendlocha},\ and\ \citenamefont
  {Klimczuk}}]{LiGa2Ir-Gornicka-exp-theory-SR}%
  \BibitemOpen
  \bibfield  {author} {\bibinfo {author} {\bibfnamefont {K.}~\bibnamefont
  {Górnicka}}, \bibinfo {author} {\bibfnamefont {G.}~\bibnamefont
  {Kuderowicz}}, \bibinfo {author} {\bibfnamefont {M.~J.}\ \bibnamefont
  {Winiarski}}, \bibinfo {author} {\bibfnamefont {B.}~\bibnamefont
  {Wiendlocha}}, \ and\ \bibinfo {author} {\bibfnamefont {T.}~\bibnamefont
  {Klimczuk}},\ }\href {\doibase 10.1038/s41598-021-95944-1} {\bibfield
  {journal} {\bibinfo  {journal} {Scientific Reports}\ }\textbf {\bibinfo
  {volume} {11}} (\bibinfo {year} {2021}),\
  10.1038/s41598-021-95944-1}\BibitemShut {NoStop}%
\bibitem [{\citenamefont {Tseng}\ \emph {et~al.}(2017)\citenamefont {Tseng},
  \citenamefont {Kuo}, \citenamefont {Lee}, \citenamefont {Chen}, \citenamefont
  {Huang}, \citenamefont {Wei}, \citenamefont {Kuo},\ and\ \citenamefont
  {Lue}}]{Ru2TaAl-Tseng-exp-theory-PRB}%
  \BibitemOpen
  \bibfield  {author} {\bibinfo {author} {\bibfnamefont {C.~W.}\ \bibnamefont
  {Tseng}}, \bibinfo {author} {\bibfnamefont {C.~N.}\ \bibnamefont {Kuo}},
  \bibinfo {author} {\bibfnamefont {H.~W.}\ \bibnamefont {Lee}}, \bibinfo
  {author} {\bibfnamefont {K.~F.}\ \bibnamefont {Chen}}, \bibinfo {author}
  {\bibfnamefont {R.~C.}\ \bibnamefont {Huang}}, \bibinfo {author}
  {\bibfnamefont {C.-M.}\ \bibnamefont {Wei}}, \bibinfo {author} {\bibfnamefont
  {Y.~K.}\ \bibnamefont {Kuo}}, \ and\ \bibinfo {author} {\bibfnamefont
  {C.~S.}\ \bibnamefont {Lue}},\ }\href {\doibase 10.1103/PhysRevB.96.125106}
  {\bibfield  {journal} {\bibinfo  {journal} {Phys. Rev. B}\ }\textbf {\bibinfo
  {volume} {96}},\ \bibinfo {pages} {125106} (\bibinfo {year}
  {2017})}\BibitemShut {NoStop}%
\bibitem [{\citenamefont {Çanl}\ \emph {et~al.}(2021)\citenamefont {Çanl},
  \citenamefont {İlhan},\ and\ \citenamefont
  {Arıkan}}]{Ir2ScGa-Murat-theory-MTC}%
  \BibitemOpen
  \bibfield  {author} {\bibinfo {author} {\bibfnamefont {M.}~\bibnamefont
  {Çanl}}, \bibinfo {author} {\bibfnamefont {E.}~\bibnamefont {İlhan}}, \
  and\ \bibinfo {author} {\bibfnamefont {N.}~\bibnamefont {Arıkan}},\ }\href
  {\doibase https://doi.org/10.1016/j.mtcomm.2020.101855} {\bibfield  {journal}
  {\bibinfo  {journal} {Materials Today Communications}\ }\textbf {\bibinfo
  {volume} {26}},\ \bibinfo {pages} {101855} (\bibinfo {year}
  {2021})}\BibitemShut {NoStop}%
\bibitem [{\citenamefont {Prakash}\ and\ \citenamefont
  {Kalpana}(2021)}]{Ir2YSi-Prakash-theory-AIPA}%
  \BibitemOpen
  \bibfield  {author} {\bibinfo {author} {\bibfnamefont {R.}~\bibnamefont
  {Prakash}}\ and\ \bibinfo {author} {\bibfnamefont {G.}~\bibnamefont
  {Kalpana}},\ }\href {\doibase 10.1063/9.0000101} {\bibfield  {journal}
  {\bibinfo  {journal} {AIP Advances}\ }\textbf {\bibinfo {volume} {11}}
  (\bibinfo {year} {2021}),\ 10.1063/9.0000101}\BibitemShut {NoStop}%
\bibitem [{\citenamefont {Samia}\ \emph {et~al.}(2022)\citenamefont {Samia},
  \citenamefont {Belkharroubi}, \citenamefont {Ibrahim}, \citenamefont {Lamia},
  \citenamefont {Saim}, \citenamefont {Maizia}, \citenamefont {Mohammed},\ and\
  \citenamefont {Al-Douri}}]{Ir2LuSb-Samia-theory-EM}%
  \BibitemOpen
  \bibfield  {author} {\bibinfo {author} {\bibfnamefont {L.}~\bibnamefont
  {Samia}}, \bibinfo {author} {\bibfnamefont {F.}~\bibnamefont {Belkharroubi}},
  \bibinfo {author} {\bibfnamefont {A.}~\bibnamefont {Ibrahim}}, \bibinfo
  {author} {\bibfnamefont {B.~F.}\ \bibnamefont {Lamia}}, \bibinfo {author}
  {\bibfnamefont {A.}~\bibnamefont {Saim}}, \bibinfo {author} {\bibfnamefont
  {A.}~\bibnamefont {Maizia}}, \bibinfo {author} {\bibfnamefont
  {A.}~\bibnamefont {Mohammed}}, \ and\ \bibinfo {author} {\bibfnamefont
  {Y.}~\bibnamefont {Al-Douri}},\ }\href {\doibase 10.1007/s42247-022-00374-y}
  {\bibfield  {journal} {\bibinfo  {journal} {Emergent Materials}\ }\textbf
  {\bibinfo {volume} {5}},\ \bibinfo {pages} {537 – 551} (\bibinfo {year}
  {2022})}\BibitemShut {NoStop}%
\bibitem [{\citenamefont {Arikan}\ \emph {et~al.}(2020)\citenamefont {Arikan},
  \citenamefont {Ocak}, \citenamefont {Dikici~Yıldız}, \citenamefont
  {Yıldız},\ and\ \citenamefont {Ünal}}]{Ir2ScAl-Arikan-theory-JKPS}%
  \BibitemOpen
  \bibfield  {author} {\bibinfo {author} {\bibfnamefont {N.}~\bibnamefont
  {Arikan}}, \bibinfo {author} {\bibfnamefont {H.~Y.}\ \bibnamefont {Ocak}},
  \bibinfo {author} {\bibfnamefont {G.}~\bibnamefont {Dikici~Yıldız}},
  \bibinfo {author} {\bibfnamefont {Y.~G.}\ \bibnamefont {Yıldız}}, \ and\
  \bibinfo {author} {\bibfnamefont {R.}~\bibnamefont {Ünal}},\ }\href
  {\doibase 10.3938/jkps.76.916} {\bibfield  {journal} {\bibinfo  {journal}
  {Journal of the Korean Physical Society}\ }\textbf {\bibinfo {volume} {76}},\
  \bibinfo {pages} {916 – 922} (\bibinfo {year} {2020})}\BibitemShut
  {NoStop}%
\bibitem [{\citenamefont {Muhammad}\ \emph {et~al.}(2019)\citenamefont
  {Muhammad}, \citenamefont {Zhang}, \citenamefont {Ali}, \citenamefont
  {Rehman},\ and\ \citenamefont {Muhammad}}]{TiZrIrZ-Muhammad-theory-TSF}%
  \BibitemOpen
  \bibfield  {author} {\bibinfo {author} {\bibfnamefont {I.}~\bibnamefont
  {Muhammad}}, \bibinfo {author} {\bibfnamefont {J.-M.}\ \bibnamefont {Zhang}},
  \bibinfo {author} {\bibfnamefont {A.}~\bibnamefont {Ali}}, \bibinfo {author}
  {\bibfnamefont {M.~U.}\ \bibnamefont {Rehman}}, \ and\ \bibinfo {author}
  {\bibfnamefont {S.}~\bibnamefont {Muhammad}},\ }\href {\doibase
  10.1016/j.tsf.2019.137564} {\bibfield  {journal} {\bibinfo  {journal} {Thin
  Solid Films}\ }\textbf {\bibinfo {volume} {690}} (\bibinfo {year} {2019}),\
  10.1016/j.tsf.2019.137564}\BibitemShut {NoStop}%
\bibitem [{\citenamefont {Hoat}(2019)}]{CoCrIrSi-Hoat-theory-CP}%
  \BibitemOpen
  \bibfield  {author} {\bibinfo {author} {\bibfnamefont {D.}~\bibnamefont
  {Hoat}},\ }\href {\doibase 10.1016/j.chemphys.2019.04.009} {\bibfield
  {journal} {\bibinfo  {journal} {Chemical Physics}\ }\textbf {\bibinfo
  {volume} {523}},\ \bibinfo {pages} {130 – 137} (\bibinfo {year}
  {2019})}\BibitemShut {NoStop}%
\bibitem [{\citenamefont {Han}\ \emph {et~al.}(2018)\citenamefont {Han},
  \citenamefont {Wu}, \citenamefont {Kuang}, \citenamefont {Yang},
  \citenamefont {Chen},\ and\ \citenamefont {Wang}}]{ZnCdIrMn-Han-theory-RP}%
  \BibitemOpen
  \bibfield  {author} {\bibinfo {author} {\bibfnamefont {Y.}~\bibnamefont
  {Han}}, \bibinfo {author} {\bibfnamefont {M.}~\bibnamefont {Wu}}, \bibinfo
  {author} {\bibfnamefont {M.}~\bibnamefont {Kuang}}, \bibinfo {author}
  {\bibfnamefont {T.}~\bibnamefont {Yang}}, \bibinfo {author} {\bibfnamefont
  {X.}~\bibnamefont {Chen}}, \ and\ \bibinfo {author} {\bibfnamefont
  {X.}~\bibnamefont {Wang}},\ }\href {\doibase 10.1016/j.rinp.2018.11.024}
  {\bibfield  {journal} {\bibinfo  {journal} {Results in Physics}\ }\textbf
  {\bibinfo {volume} {11}},\ \bibinfo {pages} {1134 – 1141} (\bibinfo {year}
  {2018})}\BibitemShut {NoStop}%
\bibitem [{\citenamefont {Han}\ \emph {et~al.}(2019)\citenamefont {Han},
  \citenamefont {Bouhemadou}, \citenamefont {Khenata}, \citenamefont {Cheng},
  \citenamefont {Yang},\ and\ \citenamefont {Wang}}]{Zn2IrMn-Han-theory-JMMM}%
  \BibitemOpen
  \bibfield  {author} {\bibinfo {author} {\bibfnamefont {Y.}~\bibnamefont
  {Han}}, \bibinfo {author} {\bibfnamefont {A.}~\bibnamefont {Bouhemadou}},
  \bibinfo {author} {\bibfnamefont {R.}~\bibnamefont {Khenata}}, \bibinfo
  {author} {\bibfnamefont {Z.}~\bibnamefont {Cheng}}, \bibinfo {author}
  {\bibfnamefont {T.}~\bibnamefont {Yang}}, \ and\ \bibinfo {author}
  {\bibfnamefont {X.}~\bibnamefont {Wang}},\ }\href {\doibase
  10.1016/j.jmmm.2018.09.053} {\bibfield  {journal} {\bibinfo  {journal}
  {Journal of Magnetism and Magnetic Materials}\ }\textbf {\bibinfo {volume}
  {471}},\ \bibinfo {pages} {49 – 55} (\bibinfo {year} {2019})}\BibitemShut
  {NoStop}%
\bibitem [{\citenamefont {Monma}\ \emph {et~al.}(2021)\citenamefont {Monma},
  \citenamefont {Roy}, \citenamefont {Suzuki}, \citenamefont {Tsuchiya},
  \citenamefont {Tsujikawa}, \citenamefont {Mizukami},\ and\ \citenamefont
  {Shirai}}]{CoIrMnAl-Monma-theory-JAC}%
  \BibitemOpen
  \bibfield  {author} {\bibinfo {author} {\bibfnamefont {R.}~\bibnamefont
  {Monma}}, \bibinfo {author} {\bibfnamefont {T.}~\bibnamefont {Roy}}, \bibinfo
  {author} {\bibfnamefont {K.}~\bibnamefont {Suzuki}}, \bibinfo {author}
  {\bibfnamefont {T.}~\bibnamefont {Tsuchiya}}, \bibinfo {author}
  {\bibfnamefont {M.}~\bibnamefont {Tsujikawa}}, \bibinfo {author}
  {\bibfnamefont {S.}~\bibnamefont {Mizukami}}, \ and\ \bibinfo {author}
  {\bibfnamefont {M.}~\bibnamefont {Shirai}},\ }\href {\doibase
  10.1016/j.jallcom.2021.159175} {\bibfield  {journal} {\bibinfo  {journal}
  {Journal of Alloys and Compounds}\ }\textbf {\bibinfo {volume} {868}}
  (\bibinfo {year} {2021}),\ 10.1016/j.jallcom.2021.159175}\BibitemShut
  {NoStop}%
\bibitem [{\citenamefont {Khandy}\ and\ \citenamefont
  {Chai}(2021)}]{Ru2TaGa-Khandy-theory-JPCS}%
  \BibitemOpen
  \bibfield  {author} {\bibinfo {author} {\bibfnamefont {S.~A.}\ \bibnamefont
  {Khandy}}\ and\ \bibinfo {author} {\bibfnamefont {J.-D.}\ \bibnamefont
  {Chai}},\ }\href {\doibase 10.1016/j.jpcs.2021.110098} {\bibfield  {journal}
  {\bibinfo  {journal} {Journal of Physics and Chemistry of Solids}\ }\textbf
  {\bibinfo {volume} {154}} (\bibinfo {year} {2021}),\
  10.1016/j.jpcs.2021.110098}\BibitemShut {NoStop}%
\bibitem [{\citenamefont {Wollmann}\ \emph {et~al.}(2015)\citenamefont
  {Wollmann}, \citenamefont {Chadov}, \citenamefont {K\"ubler},\ and\
  \citenamefont {Felser}}]{Mn2YZ-Wollmann-theory-PRB}%
  \BibitemOpen
  \bibfield  {author} {\bibinfo {author} {\bibfnamefont {L.}~\bibnamefont
  {Wollmann}}, \bibinfo {author} {\bibfnamefont {S.}~\bibnamefont {Chadov}},
  \bibinfo {author} {\bibfnamefont {J.}~\bibnamefont {K\"ubler}}, \ and\
  \bibinfo {author} {\bibfnamefont {C.}~\bibnamefont {Felser}},\ }\href
  {\doibase 10.1103/PhysRevB.92.064417} {\bibfield  {journal} {\bibinfo
  {journal} {Phys. Rev. B}\ }\textbf {\bibinfo {volume} {92}},\ \bibinfo
  {pages} {064417} (\bibinfo {year} {2015})}\BibitemShut {NoStop}%
\bibitem [{\citenamefont {Hellal}\ \emph {et~al.}(2017)\citenamefont {Hellal},
  \citenamefont {Bensaid}, \citenamefont {Doumi}, \citenamefont {Mohammedi},
  \citenamefont {Benzoudji}, \citenamefont {Azzaz},\ and\ \citenamefont
  {Ameri}}]{Mn2YGa-Tayeb-theory-CJP}%
  \BibitemOpen
  \bibfield  {author} {\bibinfo {author} {\bibfnamefont {T.}~\bibnamefont
  {Hellal}}, \bibinfo {author} {\bibfnamefont {D.}~\bibnamefont {Bensaid}},
  \bibinfo {author} {\bibfnamefont {B.}~\bibnamefont {Doumi}}, \bibinfo
  {author} {\bibfnamefont {A.}~\bibnamefont {Mohammedi}}, \bibinfo {author}
  {\bibfnamefont {F.}~\bibnamefont {Benzoudji}}, \bibinfo {author}
  {\bibfnamefont {Y.}~\bibnamefont {Azzaz}}, \ and\ \bibinfo {author}
  {\bibfnamefont {M.}~\bibnamefont {Ameri}},\ }\href {\doibase
  https://doi.org/10.1016/j.cjph.2017.01.008} {\bibfield  {journal} {\bibinfo
  {journal} {Chinese Journal of Physics}\ }\textbf {\bibinfo {volume} {55}},\
  \bibinfo {pages} {806} (\bibinfo {year} {2017})}\BibitemShut {NoStop}%
\bibitem [{\citenamefont {Fan}\ \emph {et~al.}(2020)\citenamefont {Fan},
  \citenamefont {Chen}, \citenamefont {mei Li}, \citenamefont {Hou},
  \citenamefont {Zhu}, \citenamefont {lei Luo},\ and\ \citenamefont
  {Chen}}]{Mn2YGa-LiFan-theory-JMMM}%
  \BibitemOpen
  \bibfield  {author} {\bibinfo {author} {\bibfnamefont {L.}~\bibnamefont
  {Fan}}, \bibinfo {author} {\bibfnamefont {F.}~\bibnamefont {Chen}}, \bibinfo
  {author} {\bibfnamefont {C.}~\bibnamefont {mei Li}}, \bibinfo {author}
  {\bibfnamefont {X.}~\bibnamefont {Hou}}, \bibinfo {author} {\bibfnamefont
  {X.}~\bibnamefont {Zhu}}, \bibinfo {author} {\bibfnamefont {J.}~\bibnamefont
  {lei Luo}}, \ and\ \bibinfo {author} {\bibfnamefont {Z.-Q.}\ \bibnamefont
  {Chen}},\ }\href {\doibase https://doi.org/10.1016/j.jmmm.2019.166060}
  {\bibfield  {journal} {\bibinfo  {journal} {Journal of Magnetism and Magnetic
  Materials}\ }\textbf {\bibinfo {volume} {497}},\ \bibinfo {pages} {166060}
  (\bibinfo {year} {2020})}\BibitemShut {NoStop}%
\bibitem [{\citenamefont {Nayak}\ \emph {et~al.}(2013)\citenamefont {Nayak},
  \citenamefont {Nicklas}, \citenamefont {Chadov}, \citenamefont {Shekhar},
  \citenamefont {Skourski}, \citenamefont {Winterlik},\ and\ \citenamefont
  {Felser}}]{Mn2PtGa-Nayak-exp-PRL}%
  \BibitemOpen
  \bibfield  {author} {\bibinfo {author} {\bibfnamefont {A.~K.}\ \bibnamefont
  {Nayak}}, \bibinfo {author} {\bibfnamefont {M.}~\bibnamefont {Nicklas}},
  \bibinfo {author} {\bibfnamefont {S.}~\bibnamefont {Chadov}}, \bibinfo
  {author} {\bibfnamefont {C.}~\bibnamefont {Shekhar}}, \bibinfo {author}
  {\bibfnamefont {Y.}~\bibnamefont {Skourski}}, \bibinfo {author}
  {\bibfnamefont {J.}~\bibnamefont {Winterlik}}, \ and\ \bibinfo {author}
  {\bibfnamefont {C.}~\bibnamefont {Felser}},\ }\href {\doibase
  10.1103/PhysRevLett.110.127204} {\bibfield  {journal} {\bibinfo  {journal}
  {Phys. Rev. Lett.}\ }\textbf {\bibinfo {volume} {110}},\ \bibinfo {pages}
  {127204} (\bibinfo {year} {2013})}\BibitemShut {NoStop}%
\bibitem [{\citenamefont {Nayak}\ \emph {et~al.}(2017)\citenamefont {Nayak},
  \citenamefont {Kumar}, \citenamefont {Ma}, \citenamefont {Werner},
  \citenamefont {Pippel}, \citenamefont {Sahoo}, \citenamefont {Damay},
  \citenamefont {Rößler}, \citenamefont {Felser},\ and\ \citenamefont
  {Parkin}}]{Mn2PtSn-Nayak-exp-Nature}%
  \BibitemOpen
  \bibfield  {author} {\bibinfo {author} {\bibfnamefont {A.}~\bibnamefont
  {Nayak}}, \bibinfo {author} {\bibfnamefont {V.}~\bibnamefont {Kumar}},
  \bibinfo {author} {\bibfnamefont {T.}~\bibnamefont {Ma}}, \bibinfo {author}
  {\bibfnamefont {P.}~\bibnamefont {Werner}}, \bibinfo {author} {\bibfnamefont
  {E.}~\bibnamefont {Pippel}}, \bibinfo {author} {\bibfnamefont
  {R.}~\bibnamefont {Sahoo}}, \bibinfo {author} {\bibfnamefont
  {F.}~\bibnamefont {Damay}}, \bibinfo {author} {\bibfnamefont
  {U.}~\bibnamefont {Rößler}}, \bibinfo {author} {\bibfnamefont
  {C.}~\bibnamefont {Felser}}, \ and\ \bibinfo {author} {\bibfnamefont
  {S.}~\bibnamefont {Parkin}},\ }\href {\doibase 10.1038/nature23466}
  {\bibfield  {journal} {\bibinfo  {journal} {Nature}\ }\textbf {\bibinfo
  {volume} {548}},\ \bibinfo {pages} {561} (\bibinfo {year}
  {2017})}\BibitemShut {NoStop}%
\bibitem [{\citenamefont {Meshcheriakova}\ \emph {et~al.}(2014)\citenamefont
  {Meshcheriakova}, \citenamefont {Chadov}, \citenamefont {Nayak},
  \citenamefont {R\"o\ss{}ler}, \citenamefont {K\"ubler}, \citenamefont
  {Andr\'e}, \citenamefont {Tsirlin}, \citenamefont {Kiss}, \citenamefont
  {Hausdorf}, \citenamefont {Kalache}, \citenamefont {Schnelle}, \citenamefont
  {Nicklas},\ and\ \citenamefont {Felser}}]{Mn2RhSn-Meshcheriakova-exp-PRL}%
  \BibitemOpen
  \bibfield  {author} {\bibinfo {author} {\bibfnamefont {O.}~\bibnamefont
  {Meshcheriakova}}, \bibinfo {author} {\bibfnamefont {S.}~\bibnamefont
  {Chadov}}, \bibinfo {author} {\bibfnamefont {A.~K.}\ \bibnamefont {Nayak}},
  \bibinfo {author} {\bibfnamefont {U.~K.}\ \bibnamefont {R\"o\ss{}ler}},
  \bibinfo {author} {\bibfnamefont {J.}~\bibnamefont {K\"ubler}}, \bibinfo
  {author} {\bibfnamefont {G.}~\bibnamefont {Andr\'e}}, \bibinfo {author}
  {\bibfnamefont {A.~A.}\ \bibnamefont {Tsirlin}}, \bibinfo {author}
  {\bibfnamefont {J.}~\bibnamefont {Kiss}}, \bibinfo {author} {\bibfnamefont
  {S.}~\bibnamefont {Hausdorf}}, \bibinfo {author} {\bibfnamefont
  {A.}~\bibnamefont {Kalache}}, \bibinfo {author} {\bibfnamefont
  {W.}~\bibnamefont {Schnelle}}, \bibinfo {author} {\bibfnamefont
  {M.}~\bibnamefont {Nicklas}}, \ and\ \bibinfo {author} {\bibfnamefont
  {C.}~\bibnamefont {Felser}},\ }\href {\doibase
  10.1103/PhysRevLett.113.087203} {\bibfield  {journal} {\bibinfo  {journal}
  {Phys. Rev. Lett.}\ }\textbf {\bibinfo {volume} {113}},\ \bibinfo {pages}
  {087203} (\bibinfo {year} {2014})}\BibitemShut {NoStop}%
\bibitem [{sm(2023)}]{sm}%
  \BibitemOpen
  \href@noop {} {\bibfield  {journal} {\bibinfo  {journal} {See Supplementary
  Material at [URL] for auxiliary informations about the experimental and
  computational methods, AC susceptibility, TEM and specific head data and
  their corresponding analysis and few additional simulated results.}\ }
  (\bibinfo {year} {2023})}\BibitemShut {NoStop}%
\bibitem [{\citenamefont {Rodríguez-Carvajal}(1993)}]{FullProf-suite-cite}%
  \BibitemOpen
  \bibfield  {author} {\bibinfo {author} {\bibfnamefont {J.}~\bibnamefont
  {Rodríguez-Carvajal}},\ }\href {\doibase
  https://doi.org/10.1016/0921-4526(93)90108-I} {\bibfield  {journal} {\bibinfo
   {journal} {Physica B: Condensed Matter}\ }\textbf {\bibinfo {volume}
  {192}},\ \bibinfo {pages} {55} (\bibinfo {year} {1993})}\BibitemShut
  {NoStop}%
\bibitem [{\citenamefont {Johannes~et al.}(2019)}]{PhysRevB.99.174410}%
  \BibitemOpen
  \bibfield  {author} {\bibinfo {author} {\bibfnamefont {K.}~\bibnamefont
  {Johannes~et al.}},\ }\href {\doibase 10.1103/PhysRevB.99.174410} {\bibfield
  {journal} {\bibinfo  {journal} {Phys. Rev. B}\ }\textbf {\bibinfo {volume}
  {99}},\ \bibinfo {pages} {174410} (\bibinfo {year} {2019})}\BibitemShut
  {NoStop}%
\bibitem [{\citenamefont {Simonson}\ \emph {et~al.}(2011)\citenamefont
  {Simonson}, \citenamefont {Wu}, \citenamefont {Xie}, \citenamefont {Tritt},\
  and\ \citenamefont {Poon}}]{FEM-Simonson-PRB}%
  \BibitemOpen
  \bibfield  {author} {\bibinfo {author} {\bibfnamefont {J.~W.}\ \bibnamefont
  {Simonson}}, \bibinfo {author} {\bibfnamefont {D.}~\bibnamefont {Wu}},
  \bibinfo {author} {\bibfnamefont {W.~J.}\ \bibnamefont {Xie}}, \bibinfo
  {author} {\bibfnamefont {T.~M.}\ \bibnamefont {Tritt}}, \ and\ \bibinfo
  {author} {\bibfnamefont {S.~J.}\ \bibnamefont {Poon}},\ }\href {\doibase
  10.1103/PhysRevB.83.235211} {\bibfield  {journal} {\bibinfo  {journal} {Phys.
  Rev. B}\ }\textbf {\bibinfo {volume} {83}},\ \bibinfo {pages} {235211}
  (\bibinfo {year} {2011})}\BibitemShut {NoStop}%
\bibitem [{\citenamefont {Singh}\ \emph {et~al.}(2012)\citenamefont {Singh},
  \citenamefont {Rawat}, \citenamefont {Muthu}, \citenamefont {D'Souza},
  \citenamefont {Suard}, \citenamefont {Senyshyn}, \citenamefont {Banik},
  \citenamefont {Rajput}, \citenamefont {Bhardwaj}, \citenamefont {Awasthi},
  \citenamefont {Ranjan}, \citenamefont {Arumugam}, \citenamefont {Schlagel},
  \citenamefont {Lograsso}, \citenamefont {Chakrabarti},\ and\ \citenamefont
  {Barman}}]{PhysRevLett.109.246601}%
  \BibitemOpen
  \bibfield  {author} {\bibinfo {author} {\bibfnamefont {S.}~\bibnamefont
  {Singh}}, \bibinfo {author} {\bibfnamefont {R.}~\bibnamefont {Rawat}},
  \bibinfo {author} {\bibfnamefont {S.~E.}\ \bibnamefont {Muthu}}, \bibinfo
  {author} {\bibfnamefont {S.~W.}\ \bibnamefont {D'Souza}}, \bibinfo {author}
  {\bibfnamefont {E.}~\bibnamefont {Suard}}, \bibinfo {author} {\bibfnamefont
  {A.}~\bibnamefont {Senyshyn}}, \bibinfo {author} {\bibfnamefont
  {S.}~\bibnamefont {Banik}}, \bibinfo {author} {\bibfnamefont
  {P.}~\bibnamefont {Rajput}}, \bibinfo {author} {\bibfnamefont
  {S.}~\bibnamefont {Bhardwaj}}, \bibinfo {author} {\bibfnamefont {A.~M.}\
  \bibnamefont {Awasthi}}, \bibinfo {author} {\bibfnamefont {R.}~\bibnamefont
  {Ranjan}}, \bibinfo {author} {\bibfnamefont {S.}~\bibnamefont {Arumugam}},
  \bibinfo {author} {\bibfnamefont {D.~L.}\ \bibnamefont {Schlagel}}, \bibinfo
  {author} {\bibfnamefont {T.~A.}\ \bibnamefont {Lograsso}}, \bibinfo {author}
  {\bibfnamefont {A.}~\bibnamefont {Chakrabarti}}, \ and\ \bibinfo {author}
  {\bibfnamefont {S.~R.}\ \bibnamefont {Barman}},\ }\href {\doibase
  10.1103/PhysRevLett.109.246601} {\bibfield  {journal} {\bibinfo  {journal}
  {Phys. Rev. Lett.}\ }\textbf {\bibinfo {volume} {109}},\ \bibinfo {pages}
  {246601} (\bibinfo {year} {2012})}\BibitemShut {NoStop}%
\bibitem [{\citenamefont {Venkateswara}\ \emph {et~al.}(2023)\citenamefont
  {Venkateswara}, \citenamefont {Nag}, \citenamefont {Samatham}, \citenamefont
  {Patel}, \citenamefont {Babu}, \citenamefont {Varma}, \citenamefont {Nayak},
  \citenamefont {Suresh},\ and\ \citenamefont
  {Alam}}]{YVenkateswara-FeRhCrSi-PRB}%
  \BibitemOpen
  \bibfield  {author} {\bibinfo {author} {\bibfnamefont {Y.}~\bibnamefont
  {Venkateswara}}, \bibinfo {author} {\bibfnamefont {J.}~\bibnamefont {Nag}},
  \bibinfo {author} {\bibfnamefont {S.~S.}\ \bibnamefont {Samatham}}, \bibinfo
  {author} {\bibfnamefont {A.~K.}\ \bibnamefont {Patel}}, \bibinfo {author}
  {\bibfnamefont {P.~D.}\ \bibnamefont {Babu}}, \bibinfo {author}
  {\bibfnamefont {M.~R.}\ \bibnamefont {Varma}}, \bibinfo {author}
  {\bibfnamefont {J.}~\bibnamefont {Nayak}}, \bibinfo {author} {\bibfnamefont
  {K.~G.}\ \bibnamefont {Suresh}}, \ and\ \bibinfo {author} {\bibfnamefont
  {A.}~\bibnamefont {Alam}},\ }\href {\doibase 10.1103/PhysRevB.107.L100401}
  {\bibfield  {journal} {\bibinfo  {journal} {Phys. Rev. B}\ }\textbf {\bibinfo
  {volume} {107}},\ \bibinfo {pages} {L100401} (\bibinfo {year}
  {2023})}\BibitemShut {NoStop}%
\bibitem [{\citenamefont {Kurz}\ \emph
  {et~al.}(2004{\natexlab{a}})\citenamefont {Kurz}, \citenamefont {F\"orster},
  \citenamefont {Nordstr\"om}, \citenamefont {Bihlmayer},\ and\ \citenamefont
  {Bl\"ugel}}]{FleurRef1-PRB}%
  \BibitemOpen
  \bibfield  {author} {\bibinfo {author} {\bibfnamefont {P.}~\bibnamefont
  {Kurz}}, \bibinfo {author} {\bibfnamefont {F.}~\bibnamefont {F\"orster}},
  \bibinfo {author} {\bibfnamefont {L.}~\bibnamefont {Nordstr\"om}}, \bibinfo
  {author} {\bibfnamefont {G.}~\bibnamefont {Bihlmayer}}, \ and\ \bibinfo
  {author} {\bibfnamefont {S.}~\bibnamefont {Bl\"ugel}},\ }\href {\doibase
  10.1103/PhysRevB.69.024415} {\bibfield  {journal} {\bibinfo  {journal} {Phys.
  Rev. B}\ }\textbf {\bibinfo {volume} {69}},\ \bibinfo {pages} {024415}
  (\bibinfo {year} {2004}{\natexlab{a}})}\BibitemShut {NoStop}%
\bibitem [{\citenamefont {Freimuth}\ \emph {et~al.}(2008)\citenamefont
  {Freimuth}, \citenamefont {Mokrousov}, \citenamefont {Wortmann},
  \citenamefont {Heinze},\ and\ \citenamefont {Bl\"ugel}}]{FleurRef2-PRB}%
  \BibitemOpen
  \bibfield  {author} {\bibinfo {author} {\bibfnamefont {F.}~\bibnamefont
  {Freimuth}}, \bibinfo {author} {\bibfnamefont {Y.}~\bibnamefont {Mokrousov}},
  \bibinfo {author} {\bibfnamefont {D.}~\bibnamefont {Wortmann}}, \bibinfo
  {author} {\bibfnamefont {S.}~\bibnamefont {Heinze}}, \ and\ \bibinfo {author}
  {\bibfnamefont {S.}~\bibnamefont {Bl\"ugel}},\ }\href {\doibase
  10.1103/PhysRevB.78.035120} {\bibfield  {journal} {\bibinfo  {journal} {Phys.
  Rev. B}\ }\textbf {\bibinfo {volume} {78}},\ \bibinfo {pages} {035120}
  (\bibinfo {year} {2008})}\BibitemShut {NoStop}%
\bibitem [{\citenamefont {Kurz}\ \emph
  {et~al.}(2004{\natexlab{b}})\citenamefont {Kurz}, \citenamefont {F\"orster},
  \citenamefont {Nordstr\"om}, \citenamefont {Bihlmayer},\ and\ \citenamefont
  {Bl\"ugel}}]{FleurRef3-PRB}%
  \BibitemOpen
  \bibfield  {author} {\bibinfo {author} {\bibfnamefont {P.}~\bibnamefont
  {Kurz}}, \bibinfo {author} {\bibfnamefont {F.}~\bibnamefont {F\"orster}},
  \bibinfo {author} {\bibfnamefont {L.}~\bibnamefont {Nordstr\"om}}, \bibinfo
  {author} {\bibfnamefont {G.}~\bibnamefont {Bihlmayer}}, \ and\ \bibinfo
  {author} {\bibfnamefont {S.}~\bibnamefont {Bl\"ugel}},\ }\href {\doibase
  10.1103/PhysRevB.69.024415} {\bibfield  {journal} {\bibinfo  {journal} {Phys.
  Rev. B}\ }\textbf {\bibinfo {volume} {69}},\ \bibinfo {pages} {024415}
  (\bibinfo {year} {2004}{\natexlab{b}})}\BibitemShut {NoStop}%
\bibitem [{\citenamefont {Weinert}\ \emph {et~al.}(1982)\citenamefont
  {Weinert}, \citenamefont {Wimmer},\ and\ \citenamefont
  {Freeman}}]{FleurRef4-PRB}%
  \BibitemOpen
  \bibfield  {author} {\bibinfo {author} {\bibfnamefont {M.}~\bibnamefont
  {Weinert}}, \bibinfo {author} {\bibfnamefont {E.}~\bibnamefont {Wimmer}}, \
  and\ \bibinfo {author} {\bibfnamefont {A.~J.}\ \bibnamefont {Freeman}},\
  }\href {\doibase 10.1103/PhysRevB.26.4571} {\bibfield  {journal} {\bibinfo
  {journal} {Phys. Rev. B}\ }\textbf {\bibinfo {volume} {26}},\ \bibinfo
  {pages} {4571} (\bibinfo {year} {1982})}\BibitemShut {NoStop}%
\bibitem [{\citenamefont {Graf}\ \emph
  {et~al.}(2011{\natexlab{b}})\citenamefont {Graf}, \citenamefont {Felser},\
  and\ \citenamefont {Parkin}}]{Simple-rules-TGraf-PSSC}%
  \BibitemOpen
  \bibfield  {author} {\bibinfo {author} {\bibfnamefont {T.}~\bibnamefont
  {Graf}}, \bibinfo {author} {\bibfnamefont {C.}~\bibnamefont {Felser}}, \ and\
  \bibinfo {author} {\bibfnamefont {S.~S.}\ \bibnamefont {Parkin}},\ }\href
  {\doibase https://doi.org/10.1016/j.progsolidstchem.2011.02.001} {\bibfield
  {journal} {\bibinfo  {journal} {Progress in Solid State Chemistry}\ }\textbf
  {\bibinfo {volume} {39}},\ \bibinfo {pages} {1} (\bibinfo {year}
  {2011}{\natexlab{b}})}\BibitemShut {NoStop}%
\bibitem [{foo({\natexlab{a}})}]{foot1}%
  \BibitemOpen
  \href@noop {} {\bibfield  {journal} {\bibinfo  {journal} {Specific heat of
  Mn$_2$IrGa shows a transition at/around 243 K with an estimated Sommerfeld
  parameter and the Debye temperature of $\gamma \sim$ 1.0$\pm$0.1
  mJ$\cdot$mol$^{-1}$K$^{-2}$ (comparable to that of
  metals/semimetals\cite{SpinGlass-Umetsu-exp-Metals,kittel-ssp-hc}) and
  $\theta_\mathrm{D} \sim$ 260 K respectively (see Sec. G of SM\cite{sm} for
  more details).}\ } ({\natexlab{a}})}\BibitemShut {NoStop}%
\bibitem [{foo({\natexlab{b}})}]{foot2}%
  \BibitemOpen
  \href@noop {} {\bibfield  {journal} {\bibinfo  {journal} {On positive
  $H$-axis, $\rho$ increases rapidly in a linear fashion up to 20 kOe and
  thereafter decreases for temperatures 2, 5 and 10 K (see Sec. F of
  SM\cite{sm} for more details). At 50 K, it increases linearly up to 11.5 kOe
  beyond which it becomes $H$-independent. However, for temperatures 100, 150,
  200 and 250 K, $\rho$ increases rapidly (almost linear) up to a certain field
  and then increases again (but with shallow positive slope) up to 80 kOe (see
  Sec. F of SM\cite{sm}). At 300 K (in the paramagnetic regime), a linear
  increase of $\rho$ with $H$ is attributed to the Lorentz force.}\ }
  ({\natexlab{b}})}\BibitemShut {NoStop}%
\bibitem [{\citenamefont {Park}\ \emph {et~al.}(2011)\citenamefont {Park},
  \citenamefont {Wunderlich},\ and\ \citenamefont {Marti~et al.}}]{nmat2983}%
  \BibitemOpen
  \bibfield  {author} {\bibinfo {author} {\bibfnamefont {B.}~\bibnamefont
  {Park}}, \bibinfo {author} {\bibfnamefont {J.}~\bibnamefont {Wunderlich}}, \
  and\ \bibinfo {author} {\bibfnamefont {X.}~\bibnamefont {Marti~et al.}},\
  }\href {\doibase 10.1038/nmat2983} {\bibfield  {journal} {\bibinfo  {journal}
  {Nat. Mater}\ }\textbf {\bibinfo {volume} {10}},\ \bibinfo {pages} {347}
  (\bibinfo {year} {2011})}\BibitemShut {NoStop}%
\bibitem [{\citenamefont {Cullity}\ and\ \citenamefont
  {Graham}(2011)}]{Cullity-Graham-IMM}%
  \BibitemOpen
  \bibfield  {author} {\bibinfo {author} {\bibfnamefont {B.~D.}\ \bibnamefont
  {Cullity}}\ and\ \bibinfo {author} {\bibfnamefont {C.~D.}\ \bibnamefont
  {Graham}},\ }\href@noop {} {\emph {\bibinfo {title} {Introduction to magnetic
  materials}}}\ (\bibinfo  {publisher} {John Wiley \& Sons},\ \bibinfo {year}
  {2011})\BibitemShut {NoStop}%
\bibitem [{\citenamefont {Umetsu}\ \emph {et~al.}(2017)\citenamefont {Umetsu},
  \citenamefont {Xu}, \citenamefont {Ito},\ and\ \citenamefont
  {Kainuma}}]{SpinGlass-Umetsu-exp-Metals}%
  \BibitemOpen
  \bibfield  {author} {\bibinfo {author} {\bibfnamefont {R.~Y.}\ \bibnamefont
  {Umetsu}}, \bibinfo {author} {\bibfnamefont {X.}~\bibnamefont {Xu}}, \bibinfo
  {author} {\bibfnamefont {W.}~\bibnamefont {Ito}}, \ and\ \bibinfo {author}
  {\bibfnamefont {R.}~\bibnamefont {Kainuma}},\ }\href {\doibase
  10.3390/met7100414} {\bibfield  {journal} {\bibinfo  {journal} {Metals}\
  }\textbf {\bibinfo {volume} {7}} (\bibinfo {year} {2017}),\
  10.3390/met7100414}\BibitemShut {NoStop}%
\bibitem [{\citenamefont {Kittel}(2007)}]{kittel-ssp-hc}%
  \BibitemOpen
  \bibfield  {author} {\bibinfo {author} {\bibfnamefont {C.}~\bibnamefont
  {Kittel}},\ }\href@noop {} {\emph {\bibinfo {title} {Introduction to Solid
  State Physics}}},\ \bibinfo {edition} {seventh}\ ed.\ (\bibinfo  {publisher}
  {Wiley India Pvt. Limited},\ \bibinfo {year} {2007})\BibitemShut {NoStop}%
\end{thebibliography}%

\end{document}